# Transparency versus Performance in Financial Markets: The Role of CSR Communications

*Rajiv Kashyap*
Cotsakos College of Business
William Paterson University
1600 Valley Rd.
Wayne, NJ, USA - 07470
973-720-2610
kashyapr@wpunj.edu

*Mohamed Menisy*
Cotsakos College of Business
William Paterson University
1600 Valley Rd.
Wayne, NJ, USA – 07470
973-720-2610
menisym@student.wpunj.edu

*Peter Caiazzo*
Cotsakos College of Business
William Paterson University
1600 Valley Rd.
Wayne, NJ, USA – 07470
973-720-2610
CaiazzoP@wpunj.edu

*Jim Samuel*
School of Business
University of Charleston, WV
2300 MacCorkle Ave SE
Charleston, WV 25304
973-720-2610
jim@aiknowledgecenter.com

**ABSTRACT**
Although companies are exhorted to provide more information to the financial community, it is evident that they choose different paths based upon their strategic emphasis and competitive environments. Our investigation explores the empirical boundary conditions (moderators) under which firms choose to disclose versus withhold information from investors based upon their strategic emphasis. We found significant differences in terms of voluntary information disclosures between firms that consistently delivered positive earnings surprises versus those that delivered negative earnings surprises. We investigated this effect in a more granular fashion by separately examining differences in environmental, social, and governance disclosures between the two pools of firms. We found that in essence, the differences remained consistent – and positive earnings firms were significantly more likely to disclose information about their ESG activities than their counterparts. Interestingly, none of the measures of financial performance were instrumental in distinguishing between the two pools of firms. However, our measures of reach (as measured by the number of) negative news stories lends credence to our findings. From a fund manager's perspective, this finding should raise an immediate red flag-firms that are likely to underperform are likely to be less transparent than overperformers.

## 1 CONCEPTUAL BACKGROUND

Firms utilize a variety of methods, formats, and opportunities to disclose information to the investment community. We focus upon companies' voluntary information disclosures as these offer firms with discretionary opportunities to position themselves in financial markets and more importantly, provide investors with the additional transparency they need to make more informed investment decisions. While firms utilize a variety of methods such as earnings calls, blogs, investor conferences, and so forth, to disseminate financial information about the firm's strategy and operations, their efforts to increase transparency are centered around the provision of non-financial information about their CSR and sustainability initiatives. CSR and sustainability initiatives, activities, and accomplishments are now routinely reported to the investment community along the dimensions of ESG (Environmental. Social, and Governance) activities and accomplishments. Investors can also obtain third party ratings about a company's ESG performance from agencies such as MSCI, Thomson Reuters, RobecoSAM, etc. ESG information provides the transparency needed by investors to improve risk assessment and long-term value creation (Kashyap, Mani, Basu and Caiazzo 2017). In addition to these two sources, investors can obtain information about a company's ongoing activities and strategy through newswires. In order to manage and more accurately focus our efforts, we utilized negative news stories about companies in this analysis. Negative news is more salient and has been shown to have a much greater effect on risk assessment than positive news (Dawar and Pillutla 2004). This comprises the third set of information in our analysis.

In this study, we focus upon earnings surprises, that is, the deviation of the firm's announcement of earnings from the market's consensus estimates (Pfarrer, Pollock, and Rindova 2010) for two reasons: 1) earnings surprises can occur for reasons beyond a firm's control (e.g., unexpectedly higher product sales, easing of regulatory requirements, oil spills, hurricane damage to facilities) 2) earnings surprises represent an ideal context to investigate the tradeoffs between transparency and confidentiality in financial markets. However, it is well established that firms often simultaneously engage in CSI (Corporate Social Irresponsibility) and CSR (Herzig and Moon 2013; Kang Germann and Grewal 2016; Lenz et al 2017). Despite the growing emphasis on CSR, stakeholders appear to be more heavily influenced by companies' CSI transgressions than their CSR initiatives (Kang, Germann, and Grewal 2016; Price and Sun 2017).

Earnings surprises catch investors and markets off guard resulting in abnormal (positive and negative) returns immediately after the announcements (Tucker 2007). This reaction is often contingent upon the perceived status of the company in respect to its CSR responsibilities (Kruger 2015). When companies are deemed to be socially responsible, investors often overreact to positive news and underreact to negative news. These anomalies have been explained using prospect theory, which predicts that conservative investors underreact to CSI signals (Barberis, Shleifer, and Vishny 1998). In tandem, halo effects lead investors to favor companies with a good CSR reputation (Walker, Zhang and Yu 2016). However, there is some conflicting evidence in regards to the effects of increased transparency on financial performance. On the one hand, some researchers have found that companies that are more transparent and accountable are more profitable (Eccles et al 2014). On the other, transparency has been associated with an increase in visibility leading to increased sanctions when companies default on their promises (Kruger 2015). Hence, managing CSR communications has become a top priority for managers. Note that managers strive to ensure consistency by building a reputation for delivering stakeholder value. However, when companies consistently deliver negative earnings surprises, there are strong elements of CSI and earnings management that can be inferred from such behaviors.

We began our investigation by focusing on the following questions: 1) How do companies strategically manage CSR communications to ensure a good balance between transparency and confidentiality? 2) What are the effects of increased opacity on company financial performance? CSR communications are instrumental in developing a good CSR reputation. A good CSR reputation has been viewed as a reservoir of goodwill that can buffer companies against CSI spillover effects (Groening and Kanuri 2016; Vanhamme and Grobben 2009; Kim and Woo 2019), which is often referred to as the 'insurance' effect (Kang et al 2017). In support of the insurance effect, Lenz et al (2017) found that CSR offsets the negative effects of negative events, when CSR initiatives are implemented in domains (i.e., environment, social, governance) that are different from their transgressions (e.g., a firm communicates its strong emphasis on diversity, but pays fines for an environmental violation). They attribute this to difficulties experienced by stakeholders associated with retrieving information about company activities in different domains. In addition, Kashmiri et al (2017) found that negative effects are greater for retailers that are similar to a firm that has suffered a data breach. They attribute this to a contagion effect and suggest that communications that dissociate and differentiate bystander firms from the affected firm may help reduce the spillover from negative occurrences and news.

We draw on the extant literature (e.g. Kang, Germann, and Grewal 2016; Jayachandran Kalaignanam, and Eilert 2013; Luo and Bhattacharya 2009) to develop hypotheses about the effect of ESG initiatives on financial performance. We also utilize previous research on voluntary disclosures (e.g., Giannarakis and Sariannidis 2014; Healy, Hutton, and Palepu1999; Lev 1992) to tease out the differences between the effects of ESG on earnings. We observe that investors sanction firms (Hayibor and Collins 2016) when they violate expectations of financial and ESG performance (e.g., failure to protect customer data), or equity (e.g., unfair HR practices), or threaten existing relationships (e.g. eliminate support for legacy products). Investors have been particularly quick to severely punish errant firms (Karpoff, Lee and Martin 2008; Olsen and Klaw 2017). Therefore, effectively managing the spillover effects of negative events and news on financial performance has become a top priority for managers.

Previous studies have found that negative news increases the likelihood of investor sanctions, especially when companies are deemed culpable for undesirable social outcomes (Kolbel, Busch, and Jancso 2017; Lange and Washburn 2012), and can negatively impact a firm's business associates (Kashmiri et al 2017) and MNC subsidiaries of the offending firm (Wang and Li 2019). According to Lenz et al (2017), "CSI's negative effect on firm performance may exceed CSR's positive effect…" This underscores the material and predictive implications of the effects of CSI on financial performance for stakeholders. CSI directly generates financial risk by providing stakeholders with an agenda to coordinate sanctions (Kolbel et al 2017). Increased media coverage of CSI focuses stakeholder attention and coordinates their cognitive responses to CSI occurrences. Building on this line of reasoning, we suggest that CSI invokes fear and uncertainty among stakeholders of increased regulatory scrutiny and fines, penalties, and lawsuits. CSI imposes or raises the specter of additional costs to mitigate externalities. For instance, SEC fines for insider trading and accounting fraud, FTC sanctions for false advertising, and class action lawsuits by consumers, can significantly increase company costs. This manifests in the form of increased risk for companies in capital markets as the magnitude of settlements and future compliance costs is uncertain. As a result, companies may be subject to increased costs of borrowing or fall out of favor with investors seeking stable investments.

CSI also indirectly impacts financial performance as follows. Consumer awareness of CSI is increased by media coverage (e.g., Kolbel et al 2017) and interventions by government agencies. Several government agencies protect consumer interests in the US. For instance, the Consumer Safety Product Commission (CPSC) oversees most consumer products, but food, drugs, and cosmetics are regulated by the Food and Drug Administration, (FDA), while the National Highway Traffic Safety Administration (NHTSA) watches over the automobile industry. The Securities Exchange Commission is specifically charged with protecting the financial interests of investors in US capital markets. Of particular relevance are product harm crises (Chen, Ganesan and Liu 2009; Cleeren, Dekimpe, and van Heerde 2017; Klein and Dawar 2004) and customer data breaches (Kashmiri et al 2017), which affects financial performance as well as a company's reputation and long-term prospects. Companies experience declines in revenue and simultaneously incur significant costs to repair or replace defective products. In addition, the collateral damage to brand reputation, due to negative halo effects from data breaches, has enduring and adverse implications for financial performance (Kashmiri et al 2017; Park and Sun 2017). CSI awareness in labor markets, hinders employees a company's ability to hire the best talent (Greening and Turban 2000; Jones, Willness, & Madey, 2014). In turn, this detracts from its capability to innovate and create value, thereby reducing its competitive capacity. From the above discussion, these direct and indirect effects underscore the immediate relevance and highly severe impacts of CSI on financial performance.

## 2 METHODOLOGY

We deploy an empirical approach to determining the level of transparency exhibited by firms. First, we identify companies that have consistently provided (positive +20% or negative -10) earnings surprises from 2013-2016 (16 quarters – see Figures 1 and 2 below). Consistent earnings surprises represent company attempts at earnings management and represent their strategic intent to deliberately withhold information from the investment community. This approach enabled us to identify companies that had consistently pursued an earnings management strategy. Next, we collected Bloomberg Professional's ESG disclosure scores (range: 0-100) that provide information about companies' extra financial activities for these companies. Bloomberg Professional provides disclosure based ESG scores for over 11,000 companies worldwide. ESG scores are awarded based on company filings and other publicly available information as well as a Bloomberg survey. To this data, we append information on their ESG performance using the MSCI database. The MSCI database provides performance-driven (or proprietary) ratings information on firms' environment, social, and governance performance. Finally, we use Nexis to identify all negative news about these companies during the relevant time frame.

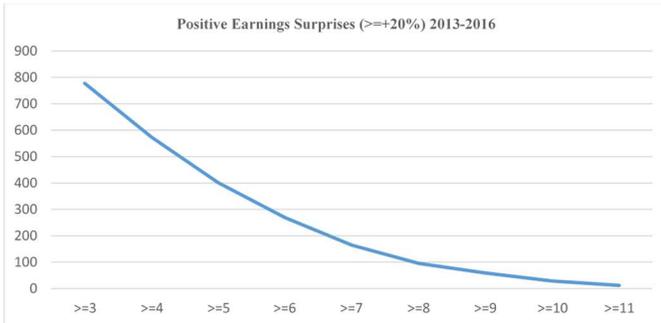
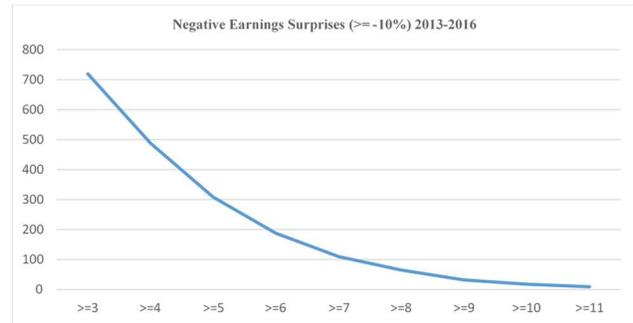

Figure 1: Positive earnings                                Figure 2: Negative earnings

A visual inspection of both plots reveals that the inflection points for both positive and negative surprises are at >=6 (i.e., when companies surprise the investor community 37.5%, that is on three or more of eight occasions). Therefore, these companies were selected for further analysis. We began our analysis with a simple comparison of means for the three sets of information identified earlier.

Table 1: Comparison of mean differences in ESG Disclosures between positive and negative earnings surprise firms.

| DIFFERENCES IN ESG DISCLOSURES | | | |
|---|---|---|---|
| | | Mean | Sig. |
| 2013ESG | 0 | 14.907299 | |
| | 1 | 18.826435 | 0.003** |
| 2014ESG | 0 | 14.618502 | |
| | 1 | 19.634613 | 0.000** |
| 2015ESG | 0 | 16.129265 | |
| | 1 | 20.808166 | 0.001** |
| 2016ESG | 0 | 16.621325 | |
| | 1 | 21.522977 | 0.001** |

** $p < 0.05$

Table 2: Comparison of mean differences in Environmental Disclosures between positive and negative earnings surprise firms

| DIFFERENCES IN ENVIRONMENTAL DISCLOSURES | | | |
|---|---|---|---|
| | | Mean | P value |
| ENV_D2013 | 0 | 14.39 | |
| | 1 | 21.66 | 0.122 |
| ENV_D2014 | 0 | 4.40 | |
| | 1 | 9.50 | .011** |
| ENV_D2015 | 0 | 4.82 | |
| | 1 | 9.36 | .011** |
| ENV_D2016 | 0 | 4.787155 | |
| | 1 | 9.647039 | .006** |

* $p < 0.1$, ** $p < 0.05$

Table 1 clearly shows that in the time periods considered for analysis (2013-2016), firms with consistently high positive earnings surprises are far more transparent than firms with negative earnings surprises in terms of their total Environmental, Social, and Governance Scores. To gain further insights, we also compared means across the three dimensions of scores. Table 2 shows that for years 2014. 2015, and 2016, the differences were in environmental disclosures were significant. In years 2014 and 2016 the differences in means of social disclosure scores were significant (Table 3). For all years 2013-16, the differences in means of governance disclosure scores were significant (Table 4). This clearly suggests that firms with positive earnings surprises are much more transparent than those with negative earnings surprises.

Table 3: Comparison of mean differences in Social Disclosures between positive and negative earnings surprise firms

| DIFFERENCES IN SOCIAL DISCLOSURES | | | |
|---|---|---|---|
| | | Mean | Sig. |
| SOC_D2013 | 0 | 14.45 | |
| | 1 | 17.90 | 0.10* |
| SOC_D2014 | 0 | 10.75 | |
| | 1 | 15.02 | 0.03** |
| SOC_D2015 | 0 | 11.98 | |
| | 1 | 14.98 | 0.07* |
| SOC_D2016 | 0 | 12.65 | |
| | 1 | 15.86 | 0.05* |

\* $p < 0.1$, \*\* $p < 0.05$,

Table 4: Comparison of mean diff. in Governance Disclosures between positive and negative earnings surprise firms (2013-16)

| DIFFERENCES IN GOVERNANCE DISCLOSURES | | | |
|---|---|---|---|
| | | Mean | Sig |
| GOV_D2013 | 0 | 50.17 | |
| | 1 | 52.41 | 0.002** |
| GOV_D2014 | 0 | 46.67 | |
| | 1 | 52.79 | 0.000** |
| GOV_D2015 | 0 | 47.52 | |
| | 1 | 53.68 | 0.000** |
| GOV_D2016 | 0 | 48.11 | |
| | 1 | 54.27 | 0.000** |

\* $p < 0.05$, \*\* $p < 0.01$,

Table 5: Comparison of mean differences in Environmental Performance between positive and negative earnings surprise firms (2013-16)

| DIFFERENCES IN ENVIRONMENTAL PERFORMANCE | | | |
|---|---|---|---|
| | GROUP | Mean | Sig. |
| Env_S2013 | 0 | 2.25 | 0.603 |
| | 1 | 2.35 | |
| Env_W2013 | 0 | 0.00 | 0.06 |
| | 1 | 0.17 | |
| Env_S2014 | 0 | 1.13 | 0.547 |
| | 1 | 1.24 | |
| Env_W2014 | 0 | 0.05 | 0.669 |
| | 1 | 0.08 | |
| Env_S2015 | 0 | 1.22 | 0.043 |
| | 1 | 1.75 | |
| Env_W2015 | 0 | 0.17 | 0.561 |
| | 1 | 0.13 | |
| Env_S2016 | 0 | 2.31 | 0.665 |
| | 1 | 2.43 | |
| Env_W2016 | 0 | 0.21 | 0.951 |
| | 1 | 0.21 | |

Table 6: Comparison of mean differences in Social Performance between positive and negative earnings surprise firms (2013-16)

| DIFFERENCES IN SOCIAL PERFORMANCE | | | |
|---|---|---|---|
| | GROUP | Mean | Sig. |
| Soc_S2013 | 0 | 3.36 | 0.745 |
| | 1 | 3.46 | |
| Soc_W2013 | 0 | 0.13 | 0.119 |
| | 1 | 0.45 | |
| Soc_S2014 | 0 | 1.72 | 0.626 |
| | 1 | 1.86 | |
| Soc_W2014 | 0 | 0.62 | 0.586 |
| | 1 | 0.73 | |
| Soc_S2015 | 0 | 1.97 | 0.425 |
| | 1 | 2.24 | |
| Soc_W2015 | 0 | 3.47 | 0.152 |
| | 1 | 3.12 | |
| Soc_S2016 | 0 | 2.68 | 0.300 |
| | 1 | 2.34 | |
| Soc_W2016 | 0 | 2.10 | 0.851 |
| | 1 | 2.03 | |

Next, we turned our attention to ESG performance scores supplied by third-party rating agencies. As the categories of performance are not identical to the disclosure categories, we focused on the overall ratings for all the firms in the sample and standardized the scores for each dimension of ESG (z- scores) to allow for comparisons across industries and categories of ESG performance. The results are shown in tables 5, 6, and 7.

Table 7: Comparison of mean differences in Governance Performance between positive and negative earnings surprise firms (2013-16)

| DIFFERENCES IN GOVERNANCE PERFORMANCE | | | |
|---|---|---|---|
| | GROUP | Mean | Sig. |
| Gov_S2013 | 0 | 0.19 | 0.117 |
| | 1 | 0.08 | |
| Gov_W2013 | 0 | 0.08 | 0.451 |
| | 1 | 0.15 | |
| Gov_S2014 | 0 | 1.72 | 0.626 |
| | 1 | 1.86 | |
| Gov_W2014 | 0 | 0.03 | 0.256 |
| | 1 | 0.08 | |
| Gov_S2015 | 0 | 0.00 | . |
| | 1 | 0.00 | |
| Gov_W2015 | 0 | 0.02 | 0.620 |
| | 1 | | |
| Gov_S2016 | 0 | 0.00 | . |
| | 1 | 0.00 | |
| Gov_W2016 | 0 | 0.33 | 0.888 |
| | 1 | 0.32 | |

Table 8: Comparison of mean differences in the number of negative news stories between positive and negative earnings surprise firms (2013-16)

| DIFFERENCES IN # NEGATIVE NEWS STORIES | | | |
|---|---|---|---|
| | | Mean | Sig |
| 2013News | 0 | 216.12 | 0.05* |
| | 1 | 106.73 | |
| 2014News | 0 | 134.02 | 0.30 |
| | 1 | 106.31 | |
| 2015News | 0 | 463.94 | 0.05* |
| | 1 | 104.87 | |
| 2016News | 0 | 359.98 | 0.06* |
| | 1 | 150.54 | |

\* p < 0.1

Finally, we focused on the third information set, that is, the number of negative news stories being disseminated through newswires. As seen from the table below, there are again substantive differences between the two categories of firms in 2013 and 2015 and the difference is marginally significant in 2016.

Table 9: Logistic Regression of Earnings Surprise Status on Information Sets and Financial Metrics (2013)

| Variable | B | S.E. | Sig. | Exp(B) | Variable | B | S.E. | Sig. | Exp(B) |
|---|---|---|---|---|---|---|---|---|---|
| INDDUMMY | 1.011 | .518 | 0.05* | 2.749 | 2013News | -.001 | .001 | 0.31 | .999 |
| ESG2013 | .043 | .022 | 0.04* | 1.044 | NETDEBT4 | -.096 | .043 | 0.02* | .908 |
| Env_S2013 | .714 | .218 | 0.00** | 2.043 | DE4 | .002 | .001 | 0.04* | 1.002 |
| Soc_S2013 | -.328 | .119 | 0.01** | .720 | ROA4 | .136 | .046 | 0.00** | 1.145 |
| Gov_S2013 | -.579 | .265 | 0.03** | .560 | Constant | -.099 | .550 | 0.86 | .906 |

\* = p<0.10, \*\* = p < 0.05

Next, we attempted to isolate the effects of these three sets of information and financial performance metrics such as risk, returns, and Tobin's Q on the propensity of firms to deliver positive and negative earnings surprises. To this end, we estimated a binary logistic model with status (0 = negative earnings surprise and 1 = positive earnings surprise) as the dependent variable and the three information sets along with financial performance and capital structure measures as the independent variables. Returns data obtained from Bloomberg and the residuals from a three-factor Fama French model Fama and French 1992, 1996) were used to compute idiosyncratic risk (Luo et al 2009) as follows:$Risk_{it}= \ln[(1-R2_{it})/ R2_{it}]$. Where the subscripts i and t refer to the specific firm and year respectively, R2 is the coefficient of determination obtained from rolling regressions of stock returns on three Fama French factors. The results are shown in Table 9 below. We mean-centered the ESG performance ratings and ESG disclosures to reduce collinearity.

As seen above, preliminary results indicate that both information disclosure (ESG2013 < 0.04), and Environmental performance (Env_S2013=-0.69, $p < 0.002$) are significant in discriminating between positive and negative earnings surprising firms. Interestingly, we also found that companies that had consistent negative earnings surprises were rated more highly than their counterparts with positive earnings surprises. We caution that these are preliminary results based upon findings from a model with data from 2013 only. In the next section, we discuss the implications of our findings.

## 3 MANAGERIAL IMPLICATIONS

We found significant differences in terms of voluntary information disclosures between firms that consistently delivered positive earnings surprises versus those that delivered negative earnings surprises. We investigated this effect in a more granular fashion by separately examining differences in environmental, social, and governance disclosures between the two pools of firms. We found that in essence, the differences remained consistent – and positive earnings firms were significantly more likely to disclose information about their ESH activities than their counterparts. From a fund manager's perspective, this finding should raise an immediate red flag – firms that are likely to underperform are likely to be less transparent than overperformers.

Interestingly, the objective data in this regard provide a less clear picture. According to third party ratings, the differences between positive surprisers and negative surprisers in terms of their ESG performance were not statistically significant. This finding in itself is extremely revealing. While managers and investors rely on ESG ratings (rather than ESG disclosures provided by companies) to predict financial performance, they may be better served by company disclosures, if they are looking for over- or under-performers. Our measures of reach (as measured by the number of) negative news stories lends more credence to our findings. There were statistical differences between the two pools, with overperformers being subject to far fewer negative news than the underperformers. Data is still being collected on this front and it is too soon to draw any concrete inferences. Additionally, none of the measures of financial performance were instrumental in distinguishing between the two pools of firms. However, we found marked differences in capital structures between the firms. Firms delivering negative earnings surprises were far more heavily leveraged than firms that delivered positive earnings surprises.

## 4 Future Research and Conclusion

This research provides interesting and novel insights into the nature of potential relationships between transparency, disclosures, corporate social responsibilities, and earnings surprises. News analytics have a behavioral component as information is reflective of behavior and behavior in financial markets can itself be driven by information, as can be revealed by textual and sentiment analytics (Samuel, 2017; Kretinin et al., 2018; Samuel et al., 2014). Similarly, alternative factors and data sources such as the role of CSR addressing bias against women, internationalization and culture, and use of Twitter data, along with data visualization strategies respectively could be used to gain richer insights (Samuel, et al., 2018, 2018; Kretinin, et al., 2019; Samuel et al., 2020, 2017; Connor, et al., 2019). In summary, our findings point to an urgent need to increase the value relevance of CSR communications to investors. A socially constructed system of voluntary disclosures can serve as a beacon of hope to investors and hopefully avert market failures. In the end, whether achieved through normative or instrumental means, an efficient market will benefit all market participants.

## 5 REFERENCES

*Barberis, Nicholas and Shleifer, Andrei and Vishny, Robert W., A Model of Investor Sentiment (February 1997). NBER Working Paper No. w5926. Available at SSRN: https://ssrn.com/abstract=225707*

*Conner, C., Samuel, J., Kretinin, A., Samuel, Y., & Nadeau, L. (2020). A picture for the words! textual visualization in big data analytics. arXiv preprint arXiv:2005.07849.*

*Dawar, N., & Pillutla, M. M. (2000). Impact of product-harm crises on brand equity: The moderating role of consumer expectations," Journal of Marketing Research, 37, 215–226*

*Du, Shuili, Kun Yu, C. B. Bhattacharya, and Sankar Sen. 2017. "The Business Case for Sustainability Reporting: Evidence from Stock Market Reactions." Journal of Public Policy & Marketing 36, no. 2: 313-330.*

*Eccles, Ioannou, and Serafeim (2013), "The Impact of Corporate Sustainability, Management Science 60(11), pp. 2835–2857.*

*Fama, E.F., & French, K.R. (1992). The cross-section of expected stock returns. Journal of Finance, 47(2), 427-65.*

*Fama, E.F., & French, K.R. (1996). Multifactor Explanations of Asset Pricing Anomalies. Journal of Finance 51(1), 55-84.*

*Giannarakis, G., Konteos, G., & Sariannidis, N. (2014). Financial, governance and environmental determinants of corporate social responsible disclosure. Management Decision, 52(10), 1928.*

*Groening, Christopher & Kanuri, Vamsi. (2013). Investor reaction to positive and negative corporate social events. Journal of Business Research. 66. 1852-1860.*

*Hayibor, S., & Collins, C. (2016). Motivators of mobilization: JBE JBE. Journal of Business Ethics, 139(2), 351-374.*

*Healy, P., Hutton, A.P. and Palepu, K.G. (1999), "Stock performance and intermediation changes surrounding sustained increases in disclosure," Contemporary Accounting Research, Vol. 16 No. 3, pp. 485-520.*

*Herzig, C., & Moon, J. (2013). Discourses on corporate social ir/responsibility in the financial sector. Journal of Business Research, 66(10), 1870.*

*Jayachandran, S., Kalaignanam, K. and Eilert, M. (2013), Product and environmental social performance: Varying effect on firm performance. Strat. Mgmt. J., 34: 1255-1264.*

*Kang, Charles, Frank Germann, and Rajdeep Grewal. "Washing Away Your Sins? Corporate Social Responsibility, Corporate Social Irresponsibility, and Firm Performance." Journal of Marketing 80.2 (2016): 59-79.*

*Karpoff, J. M., Lee, D. S., & Martin, G. S. (2008). The cost to firms of cooking the books. Journal of Financial and Quantitative Analysis, 43(3), 581.*

*Kashmiri, S., Nicol, C. D., & Hsu, L. (2017). Birds of a feather: Intra-industry spillover of the target customer data breach and the shielding role of IT, marketing, and CSR. Journal of the Academy of Marketing Science, 45(2), 208-228.*

*Klein, Jill, and Niraj Dawar. "Corporate social responsibility and consumers' attributions and brand evaluations in a product–harm crisis." International Journal of research in Marketing 21.3 (2004): 203-217.*

*Kolbel, Julian F., Timo Busch, and Leonhardt M. Jancso. "How Media Coverage of Corporate Social Irresponsibility Increases Financial Risk." Strategic Management Journal 38.11 (2017): 2266-84.*

*Lenz, I., Wetzel, H. A., & Hammerschmidt, M. (2017). Can doing good lead to doing poorly? firm value implications of CSR in the face of CSI. Journal of the Academy of Marketing Science, 45(5), 677-697.*

*Lev, Baruch, (Summer 1992), "Information Disclosure Strategy," California Management Review, 34, 4, 0-32.*

*Luo, Xueming and Bhattacharya, Chitrabhanu, The Debate Over Doing Good: Corporate Social Performance, Strategic Marketing Levers, and Firm-Idiosyncratic Risk (2009). Journal of Marketing, 73(6), 198-213, 2009. Available at SSRN: https://ssrn.com/abstract=2260348*

*Nikolaeva, R., & Bicho, M. (2011). The role of institutional and reputational factors in the voluntary adoption of corporate social responsibility reporting standards. Journal of the Academy of Marketing Science, 39(1), 136-157.*

*Kashyap, Rajiv, Sudha Mani, Sam BAsu and Peter Caiazzo. (2017). Talking the Walk: Do CSR Communications Improve Financial Performance? Working Paper, BPPRF, Cotsakos College of Business, William Paterson University*

*Kretinin, A., Samuel, J., & Kashyap, R. (2018). When the going gets tough, the tweets get going! an exploratory analysis of tweets sentiments in the stock market. American Journal of Management, 18(5).*

*Kretinin, A., Samuel, J., & Kashyap, R. (2019). Do Family Firms Prefer Global Intensity to Global Reach? An Analysis of the Role of Geographical and Cultural Distances Upon Internationalization of Family Firms. Journal of Business and Economic Studies, 23(1), 55-72.*

*Kruger P. (2015) Corporate goodness and shareholder wealth. J Finance Econ 115:304–329*

*Olsen, T. D., & Klaw, B. W. (2017). Do investors punish corporations for malfeasance?: Disclosure, materiality and market reactions to corporate irresponsibility. The Journal of Corporate Citizenship, 2017(65), 56.*

*Pfarrer, M.D., Pollock, T.G. and Rindova, V.P. (2010) A Tale of Two Assets: The Effects of Firm Reputation and Celebrity on Earnings Surprises and Investor's Reactions. Academy of Management Journal, 53, 1131-1152.*

*Price, Joseph M, and Sun, Wenbin. "Doing Good and Doing Bad: The Impact of Corporate Social Responsibility and Irresponsibility on Firm Performance." Journal of Business Research 80 (2017): 82–97.*

*Samuel, J. (2016). An Analysis of Technological Features Enabled Management of Information Facets.*

*Samuel, J. (2017). Information token driven machine learning for electronic markets: Performance effects in behavioral financial big data analytics. JISTEM-Journal of Information Systems and Technology Management, 14(3), 371-383.*

*Samuel, J., Holowczak, R., Benbunan-Fich, R., & Levine, I. (2014, January). Automating discovery of dominance in synchronous computer-mediated communication. In 2014 47th Hawaii International Conference on System Sciences (pp. 1804-1812). IEEE.*

*Samuel, Y., George, J., & Samuel, J. (2018, April). Beyond Stem, How Can Women Engage Big Data, Analytics, Robotics and Artificial Intelligence? An Exploratory Analysis Of Confidence And Educational Factors In The Emerging Technology Waves Influencing The Role Of, And Impact Upon, Women. In 2018 NEDSI Annual Conference (47th) (p. 359)*

*Samuel, J., Holowczak, R., & Pelaez, A. (2017). The Effects Of Technology Driven Information Categories On Performance In Electronic Trading Markets. Journal of Information Technology Management, 28(1-2), 1.*

*Samuel, J., Kashyap, R., & Kretinin, A. (2018). Going Where the Tweets Get Moving! An Explorative Analysis of Tweets Sentiments in the Stock Market. Proceedings of the Northeast Business & Economics Association.*

*Samuel, J. & Pelaez, A., (2017). Informatics in Information Richness: A Market Mover? An Examination of Information Richness in Electronic Markets. In the Proceedings of The 8th ICSIT, Orlando, Florida, International Institute of Informatics and Systemics.*

*Samuel, J., Ali, G. G., Rahman, M., Esawi, E., & Samuel, Y. (2020). Covid-19 public sentiment insights and machine learning for tweets classification. Information, 11(6), 314.*

*Samuel, J., Rahman, M. M., Ali, G. M. N., Samuel, Y., Pelaez, A., Chong, P. H., & Yakubov, M. (2020). Feeling Positive About Reopening? New Normal Scenarios from COVID-19 Reopen Sentiment Analytics. medRxiv. Digital Object Identifier: 10.1109/ACCESS.2020.3013933*

*Tucker, J. W. (2007). Is openness penalized? stock returns around earnings warnings. The Accounting Review, 82(4), 1055-1087.*

*Walker, K., Zhang, Z., & Yu, B.The angel-halo effect. European Business Review, 28, 709-722.*